\documentclass[pra,epsfig,twocolumn,showpacs]{revtex4}

\usepackage{graphicx}
\usepackage[latin1]{inputenc}
\usepackage{latexsym}
\usepackage{epsfig}

\begin{document}
\title{The Mott insulator phase of the one dimensional Bose-Hubbard model:
a high order perturbative study}
\author{Bogdan Damski$^1$ and Jakub Zakrzewski$^2$}
\affiliation{
(1) Theory Division, Los Alamos National Laboratory, MS-B213, Los Alamos, NM 87545, USA\\
(2) Instytut Fizyki imienia Mariana Smoluchowskiego and Mark Kac Complex
Systems Research Center, Uniwersytet Jagiello\'nski, ulica Reymonta 4,
PL-30-059 Krak\'ow, Poland}
\begin{abstract}

The one dimensional Bose-Hubbard model  at a unit filling factor is studied by means
of  a very high order symbolic perturbative expansion. 
Analytical expressions are derived for the ground state quantities such as 
energy per site, variance of on-site occupation, and correlation functions:
$\langle\hat{a}_j^\dag\hat{a}_{j+r}\rangle$ and $\langle \hat{n}_j\hat{n}_{j+r}\rangle$. 
These findings  are compared to numerics  and good agreement
is found in the Mott insulator phase. 
Our results provide  analytical approximations to important observables 
in the Mott phase, and 
are also of direct relevance  to
future experiments with  ultra cold atomic gases placed in optical  lattices.
We also discuss the symmetry of the Bose-Hubbard model associated with 
the sign change of the tunneling coupling. 
\end{abstract}
\maketitle

\section{Introduction}
One of the most fascinating recent trends in physics of cold gases
concerns atomic gases in optical lattices \cite{jaksch,bloch,esslinger}. 
These systems offer  ``atomic Hubbard toolbox'' \cite{annals} 
that can be used for studies of condensed matter models in a uniquely
controlled manner. Cold atoms in optical lattices can 
be used for investigation of high $T_c$ superconductivity \cite{hightc},
disordered systems \cite{sanpera}, various spin models \cite{duan},
novel quantum magnets \cite{bodzio1}, etc.
Perhaps the most important  version of the Hubbard model, 
that can be studied in optical lattices, is the Bose-Hubbard model (BHM) 
\cite{fisher}. 
This model is a prototypical system on which understanding of 
quantum phase transitions (QPTs) in boson systems is based \cite{sachdev}, and 
it has important applications in  construction of a quantum computer
\cite{qc}.

Despite lots of theoretical  studies on  
the BHM \cite{dmrg,qmc,krauth,white,batrouni,elstner}, there 
is  still lack of  analytical predictions about some basic
experimentally relevant quantities. This motivates us to 
present here the results of a high order perturbative expansion
in the tunneling coupling. These results, providing analytical 
approximations to different physical quantities in the Mott phase, 
should be helpful in the interpretation of experimental data. 

We focus on  a one dimensional (1D) homogeneous system  at average 
density of one atom per site, i.e., on the simplest  BHM  
undergoing a QPT. We calculate the following {\it ground state} quantities:
energy per site ${\cal E}$, atom-atom correlations
$C(r)=\langle\hat{a}_j^\dag\hat{a}_{j+r}\rangle$, density-density correlations
$D(r)=\langle \hat{n}_j\hat{n}_{j+r}\rangle$, and 
variance of on-site number operator 
${\rm Var}(\hat{n})=[\langle\hat{n}_j^2\rangle-\langle\hat{n}_j\rangle^2]^{1/2}$.

The Hamiltonian of interest, in terms of dimensionless variables
used through the paper, reads
\begin{equation}
\label{H}
\hat{H}= -J \sum_{i=1}^M(\hat{a}_{i+1}^\dag \hat{a}_i + {\rm h. c.})
+\frac{U}{2} \sum_{i=1}^M\hat{n}_i(\hat{n}_i-1),
\end{equation}
where we denote  the number of atoms, and the number of lattice sites as $M$.
Since physics of the BHM depends on the $J/U$ ratio only, we set
$U\equiv1$ for convenience. With this choice, 
the critical point  between the  Mott insulator (MI)  ($0\le J< J_c$) and 
superfluid (SF) ($J>J_c$)  is at $J_c\approx0.29$ \cite{white}.

Though the great deal of attention was recently devoted to cold atoms 
in inhomogeneous  lattices \cite{bloch,esslinger,batrouni}, the homogeneous 
systems described by the Hamiltonian (\ref{H}) can be realized in the nearest
future in at least two setups. First, there is a recent experiment done in the Raizen group
\cite{raizen}, where a single, one dimensional, homogeneous box is realized in a proper
configuration of laser beams. After superposing a 
standing laser field on it, a 1D {\it homogeneous} Bose-Hubbard model (\ref{H})
with open boundary
conditions can be achieved. Since it was already demonstrated in \cite{raizen}
that one can load this 1D box with ultracold bosons and then count  them 
 very efficiently, the studies of
lattices with a desired number of atoms per site should be available (as
discussed below on specific examples).
Another experimental opportunity shows up after realization of the
ring-shaped optical lattice proposed recently in Ref. \cite{osterloh}.
This time, a 1D {\it homogeneous} lattice with periodic boundary conditions should
be available for experimental investigations.

\section{The method}
Our  findings come from  a  high order symbolic perturbative expansion in the 
tunneling coupling. This method  was  first successfully  applied to  the 
calculation of a ground state and excited states of the BHM 
in one and two dimensional systems  by  Elstner and
Monien \cite{elstner}.
We compare perturbative expansions to  numerical data obtained using 
the imaginary time evolution with the so-called Vidal's algorithm 
\cite{Vidal}, which is equivalent to Density Matrix Renormalization 
Group scheme \cite{collath}. This allows for verification of accuracy 
of our analytical predictions. The Vidal's algorithm calculations 
assume open boundary conditions, which breaks the translational invariance 
of the system.  To minimize finite size effects during comparison between 
the numerics and perturbative expansions valid for infinite systems
(where boundary conditions are irrelevant) we have calculated
the correlation functions around the system center. 

We aim at calculation of high order perturbative corrections 
to different quantities of interest in the infinite Bose-Hubbard
model. The perturbation theory is developed around the Fock state
$|1,1,\dots\rangle$, where the numbers are boson on-site occupations.
The expansion is done in the $J$ parameter (\ref{H}).

In principle, the calculations can be performed by hand 
by  perturbative determination of the wave-function up to 
a given order in the infinite system, and then subsequent calculation
of expectation values in this wave-function. The attainable 
order of the expansion, however, is very 
limited (the wave-function can be determined up to the $J^3\sim J^4$
terms) so this method is not an option here. 

A better alternative 
is to perform a linked cluster expansion (LCE) \cite{gelfand} that 
has been used so far in spin systems \cite{spins} and 
the Bose-Hubbard model \cite{elstner}. 
This method consists of two steps. First, one has to generate 
all the clusters (sets of lattice sites in the BHM \cite{elstner})
that contribute to a given order of expansion. The largest of these 
clusters have the size comparable to the order of expansion. 
Then, one has to perform a perturbative
expansion in the relevant  clusters
and sum up the results properly. In the end, one gets perturbative
expansion of different quantities valid for infinite system from analysis
of  finite clusters. Since the whole procedure
can be implemented on a computer in a symbolic way, the high order
expansions become feasible.

The key for our calculations message from the LCE 
is the following:  all the information about $i$-th order perturbative expansions 
in the infinite system is encoded in the small subsystems of the size $\sim i$.
In accordance with this statement, we observe for 
every non-extensive 
observable that we consider,  say  $O$, that when we do a  perturbative expansion 
$O|_M=\sum_i O_i|_M J^i$ ($M$ is the system  size) the following holds 
\begin{equation}
O_i|_M\equiv O_i|_{M_c}, \ \ M\ge M_c(i),
\label{condition}
\end{equation}
i.e., the perturbative corrections become size independent for large enough 
systems. Naturally, $M_c(i)$ grows with  $i$, but
the growth is ``reasonably'' slow: $M_c(i)\sim i$ (see Table \ref{tablica}).
We can not, of course, check explicitly  the relation (\ref{condition})
for arbitrarily large $M$, but it is clear from the LCE that the size 
independent expansion terms ($O_i|_{M\ge M_c}$)  correspond to infinite system 
predictions.

In our calculations we make a direct use of the relation  (\ref{condition})
avoiding therefore generation of the cluster states. This simplifies the computer
implementation of the whole procedure, but probably leads to more stringent 
requirements on the computer resources. To be more specific,
we fix the system size $M$, assume  periodic boundary conditions 
to keep the system translationally invariant, and do a standard perturbative expansion
leading to determination of  the ground state wave function
up to a given order (Appendix \ref{pe}).
Having the ground state wave-function, we  calculate perturbative corrections
to the expectation values of different operators and study their dependence
on the system size $M$. This way we easily get $M_c(i)$  that guarantees 
(\ref{condition}): Table \ref{tablica}, as well as the desired perturbative 
corrections valid for infinite systems. Notice that due to (\ref{condition})
all finite size contributions are filtered out from the perturbative
expansions for large enough $M$.
To illustrate these findings we note that
the  2nd order correction to the ground state energy per site in 
3 sites and 3 atoms  system is the same as in the
infinite model at unit filling factor -- a result that can be easily verified analytically. 

\begin{table}
\begin{tabular}{c c c}
\hline\hline
${\cal E}$ & $C(r)$ & $D(r)$\\ \hline
$M\ge i+1$  \ \ & $M\ge i+1+r$  \ \ & $M\ge i+1$ and $2r\le i$\\ \hline\hline
\end{tabular}
\caption{Conditions on the system size $M$ for getting size independent 
perturbative predictions at $i$-th order to ${\cal E}$, $C(r)$ and $D(r)$
-- see (\ref{condition}). These results are obtained for the system filled with 
one atom per site.}
\label{tablica}
\end{table}

\section{Perturbative expansions}

To start, the ground state energy per site, ${\cal E}$, satisfies   
\begin{eqnarray}
\frac{{\cal E}}{4}&&= -J^2+J^4+\frac{68}{9}J^6-
\frac{1267}{81} J^8 + \frac{44171}{1458} J^{10}-\nonumber\\
&&\frac{4902596}{6561} J^{12} -\frac{8020902135607}{2645395200} J^{14}
+{\cal O}(J^{16}).
\label{energy_per_site}
\end{eqnarray}
Before proceeding further, 
it should be stressed  that the fractions come from a symbolic calculation,
whose details are presented in Appendix \ref{pe}.

Coming back to (\ref{energy_per_site}), we notice that so far the largest 
published order of ${\cal E}$ expansion was the sixth \cite{elstner}. Naturally,
the expansion of ${\cal E}$ from  \cite{elstner} matches first three terms
of (\ref{energy_per_site}). 
Changing fractions into numbers one gets approximately 
the following sequence  of non-zero coefficients 
$\{-1, 1, 8, -16, 30, -747, -3032\}$, which  shows that the series has a rather
unpredictable form.

\begin{figure}[t]
\includegraphics[width=\columnwidth,clip=true]{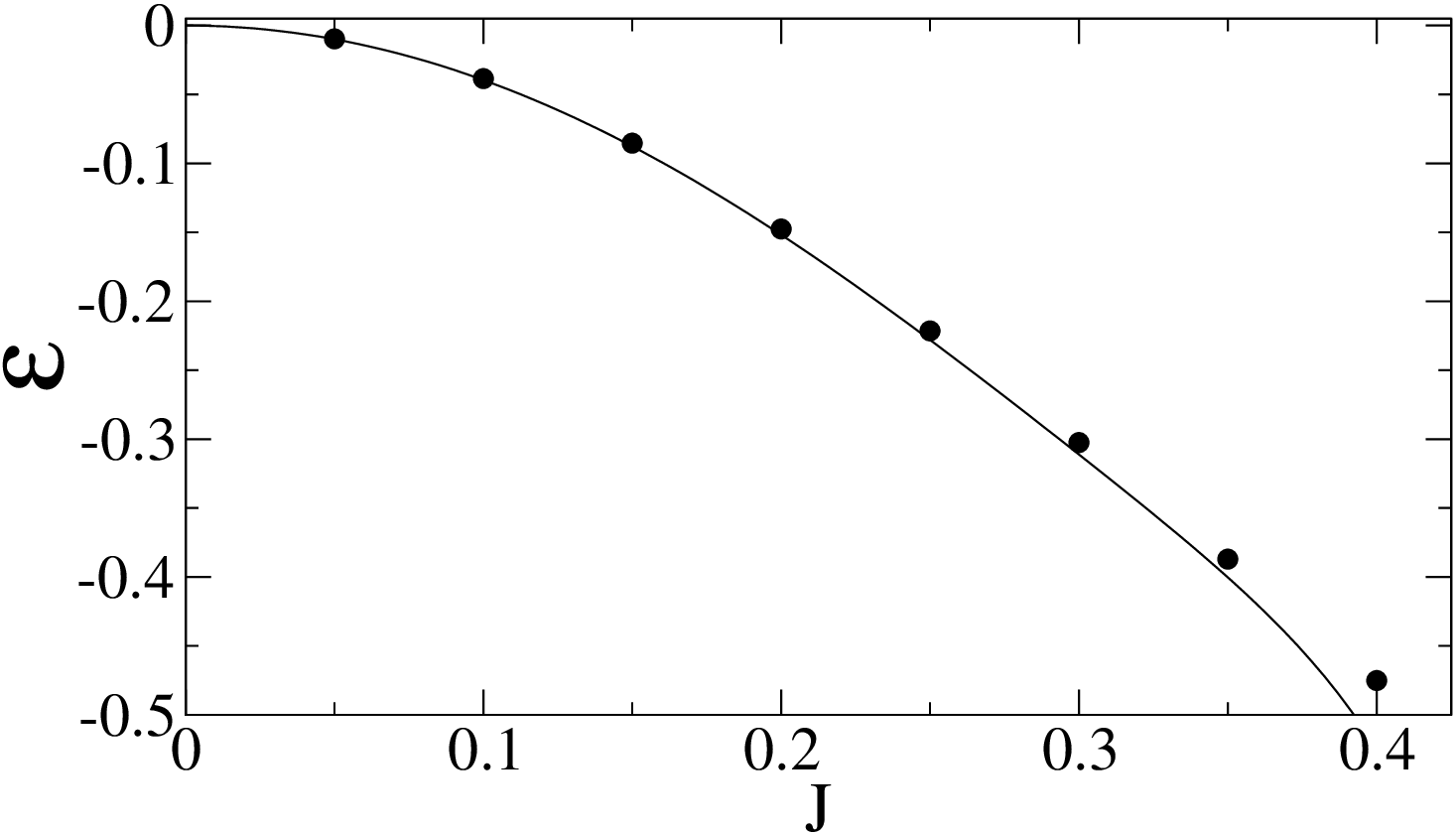}
\caption{The perturbative prediction for the ground state energy per site from 
(\ref{energy_per_site})  (solid line) vs. numerics for $M=40$ system (dots). 
}
\label{fig1}
\end{figure}

To estimate accuracy of the expansion we plot (\ref{energy_per_site}) 
vs. numerical data for a fairly large system of 
$40$ atoms in $40$ sites: Fig. \ref{fig1}. This plot shows that 
there is quite a good agreement between (\ref{energy_per_site})  and 
numerics for $J$ smaller than about $0.3$, i.e., at least in a MI phase. 
The discrepancies present in Fig. \ref{fig1} (and all  other 
figures in this paper), may come from the following  sources. 
First, our numerics is done in a finite system with open boundary 
conditions, while perturbative expansion yields results for infinite system.
Second, we might need more expansion terms  to get
more accurate  predictions. 
Third, the full perturbative expansion may fail to converge 
for large enough $J$'s,  probably $J\ge J_c$. In the worst case,
the series might be  of asymptotic kind as  was shown to be the case 
in some other systems \cite{zin}. Further discussion about 
convergence of expansions presented here is  beyond the
scope of the present contribution.

\begin{figure}[t]
\includegraphics[width=\columnwidth, clip=true]{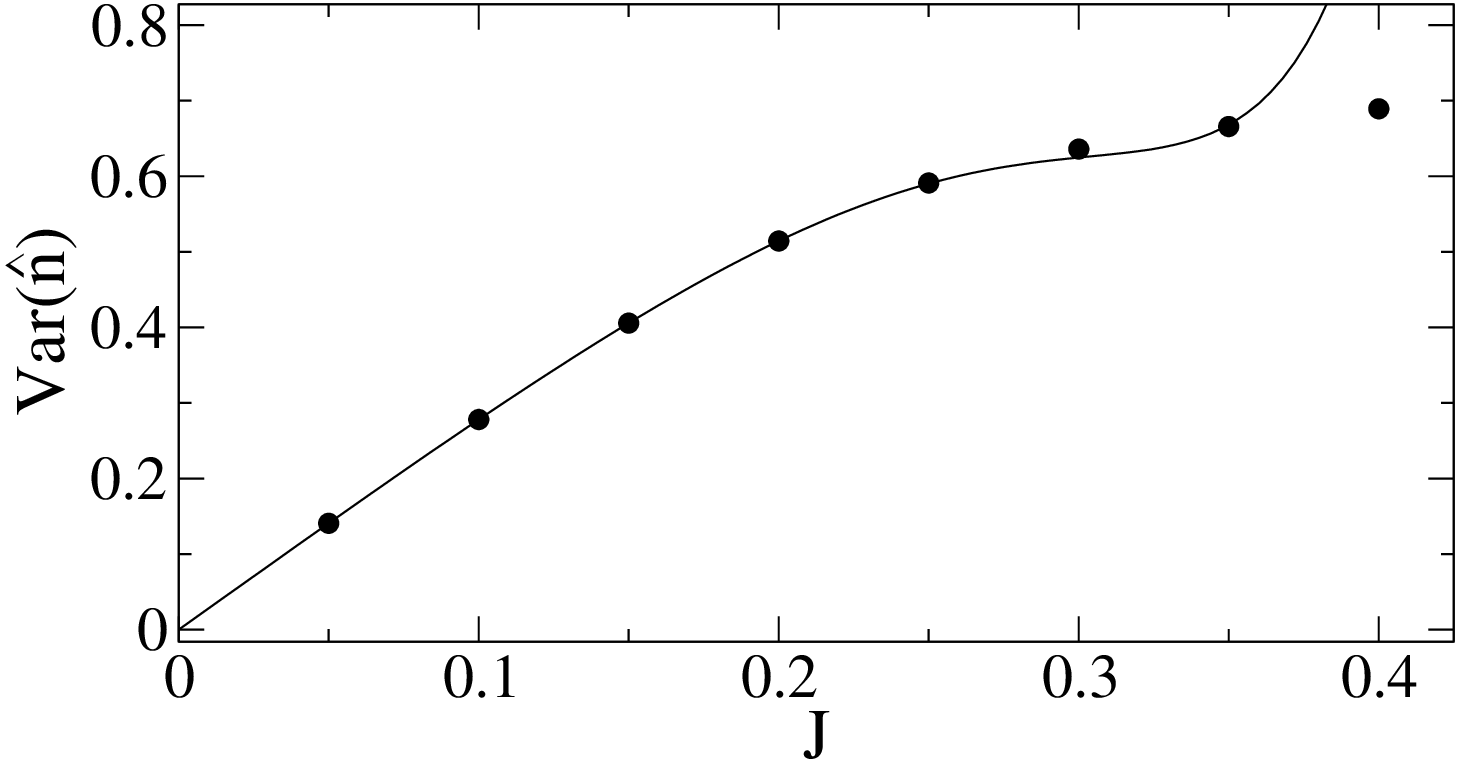}
\caption{The perturbative result for the variance of on-site number operator from (\ref{V})
         (solid line) vs. numerics for $M=40$ system (dots).}
\label{fig2}
\end{figure}

The variance of on-site number operator, ${\rm
Var}(\hat{n})$, up to  ${\cal O}(J^{15})$ terms satisfies
\begin{eqnarray}
\label{V}
\frac{{\rm Var}(\hat{n})}{\sqrt{2}}&=& 2J- 3J^3-\frac{1441}{36}J^5
+\frac{32045}{648}J^7 -\frac{3105413}{5184}J^{9}\nonumber\\ &+& 
\frac{6979423019}{839808}J^{11}+ \frac{207832615291307}{5290790400}
J^{13},
\end{eqnarray}
which is illustrated in Fig. \ref{fig2}. As for the ${\cal E}$
expansion, the agreement between perturbative prediction and numerics
is good for $J$ smaller than $\sim0.35$. Thus, our result 
can be well applied to the system in the MI phase. Most interestingly,
it shows that the variance of site occupation on the MI - SF boundary 
equals as much as $60\%$ of its deep superfluid value in the limit of
$J\to\infty$ when the system
is maximally delocalized (\ref{sf}). 

Expansion (\ref{V}) is  useful, e.g.,  because there is an 
ongoing experiment that aims at  ${\rm Var}(\hat{n})$
determination in a 1D {\it untrapped} setup \cite{raizen}.
The measurement of ${\rm Var}(\hat{n})$
can be possible due to ability for a high efficiency single
atom detection already shown in \cite{raizen}. It can be performed 
once  extraction of  atoms from a single lattice site will be
demonstrated. Extracted atoms can be counted, and then all the remaining atoms
from a lattice can be released and counted. Averaging  the results 
of single site countings  over the 
measurements where total number of atoms is  close to number of lattice sites,
one should get experimentally ${\rm Var}(\hat{n})$. 
Another aspect of this 
experiment is that the lattice is blocked at ends with 
 laser beams, which corresponds  
to open boundary conditions used in all our numerical calculations.

\begin{figure}[t]
\includegraphics[width=\columnwidth,clip=true]{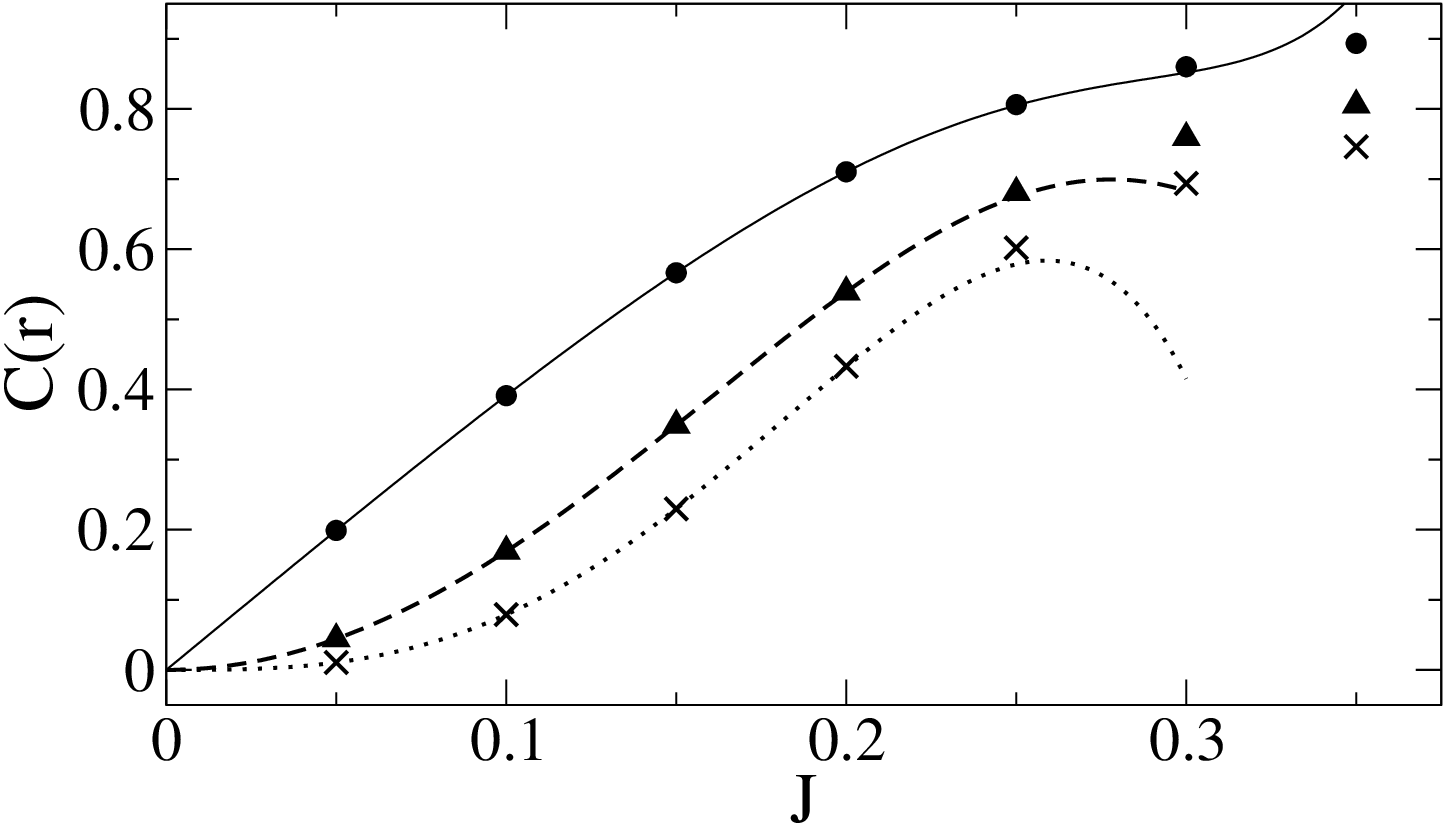}
\caption{Atom-atom correlations as a function of the tunneling $J$.
        Numerics for $M=40$ system: $C(1)$ (dots), $C(2)$ (triangles), $C(3)$ (x's).
        Lines: solid, dashed, dotted represent perturbative expansions (\ref{K})
        of $C(1)$, $C(2)$, $C(3)$, respectively.}
\label{fig3}
\end{figure}

Other quantities of interest are atom-atom correlation functions $C(r)$ defined above.
They were previously studied numerically in the one-dimensional 
Bose-Hubbard model in \cite{dmrg} and perturbatively in 
two dimensional Bose-Hubbard models in \cite{elstner}.
They are important because  another measurable quantity,
the momentum distribution of atoms in a lattice, is expressed as 
$\sim\sum_r C(r)\exp(ikr)$ \cite{zwerger}, with $k$ being atomic momentum. 
Here we list a few most important ones
\begin{eqnarray}
\label{K}
C(1) &=& 4J-8J^3
-\frac{272}{3}J^5+\frac{20272}{81}J^7
- \frac{441710}{729} J^9 \nonumber \\ &+& \frac{39220768}{2187} J^{11}
+ \frac{8020902135607}{94478400}J^{13}+   {\cal O}(J^{15}),\nonumber\\
C(2)&=& 18J^2-\frac{320}{3}J^4
-\frac{1826}{9}J^6+\frac{234862}{243}J^8\nonumber \\ &+&
\frac{345809}{2916} J^{10} + 
\frac{4434868108963}{220449600} J^{12}+
{\cal O}(J^{14}),\nonumber\\
C(3)&=& 88J^3-
\frac{8324}{9}J^5+\frac{126040}{81}J^7 
+ \frac{7883333}{486} J^9\nonumber \\ &-&
\frac{220980576341}{1049760} J^{11}+ {\cal O}(J^{13}).\nonumber\\
\end{eqnarray}
The comparison between expansions (\ref{K}) and numerics (Fig.~\ref{fig3})
shows good agreement up to $J$ equal to $0.25\sim0.3$, i.e., almost in an 
entire MI phase. Additionally,  (\ref{K})  and our 
results for $C(4\cdots7)$ indicate that $C(r>0)={\cal O}(J^r)$.

Expansions (\ref{K}) reveal  that 
atom-atom correlations  take very 
substantial values at the critical point, e.g., $C(1)$ at $J_c$ equals 
about $0.8$, i.e., $80\%$ of its deep superfluid value (\ref{sf}):
an interesting  result showing that system wave function
departs significantly from the $|1,1,\dots\rangle$ state at the 
critical point.

Finally we discuss the density-density correlations. They can be determined in a 
counting experiment almost the same as the one discussed for ${\rm Var}(\hat{n})$   
measurement, except
for the fact that now atoms from two depleted sites have to be counted first. 
A perturbation theory predicts (\ref{D}) with accuracy of ${\cal O}(J^{16})$.
This data and our results for $D(4\cdots7)$ reveal that 
$D(r>0)= 1 - {\cal O}(J^{2r})$.

\begin{figure}[t]
\includegraphics[width=\columnwidth,clip=true]{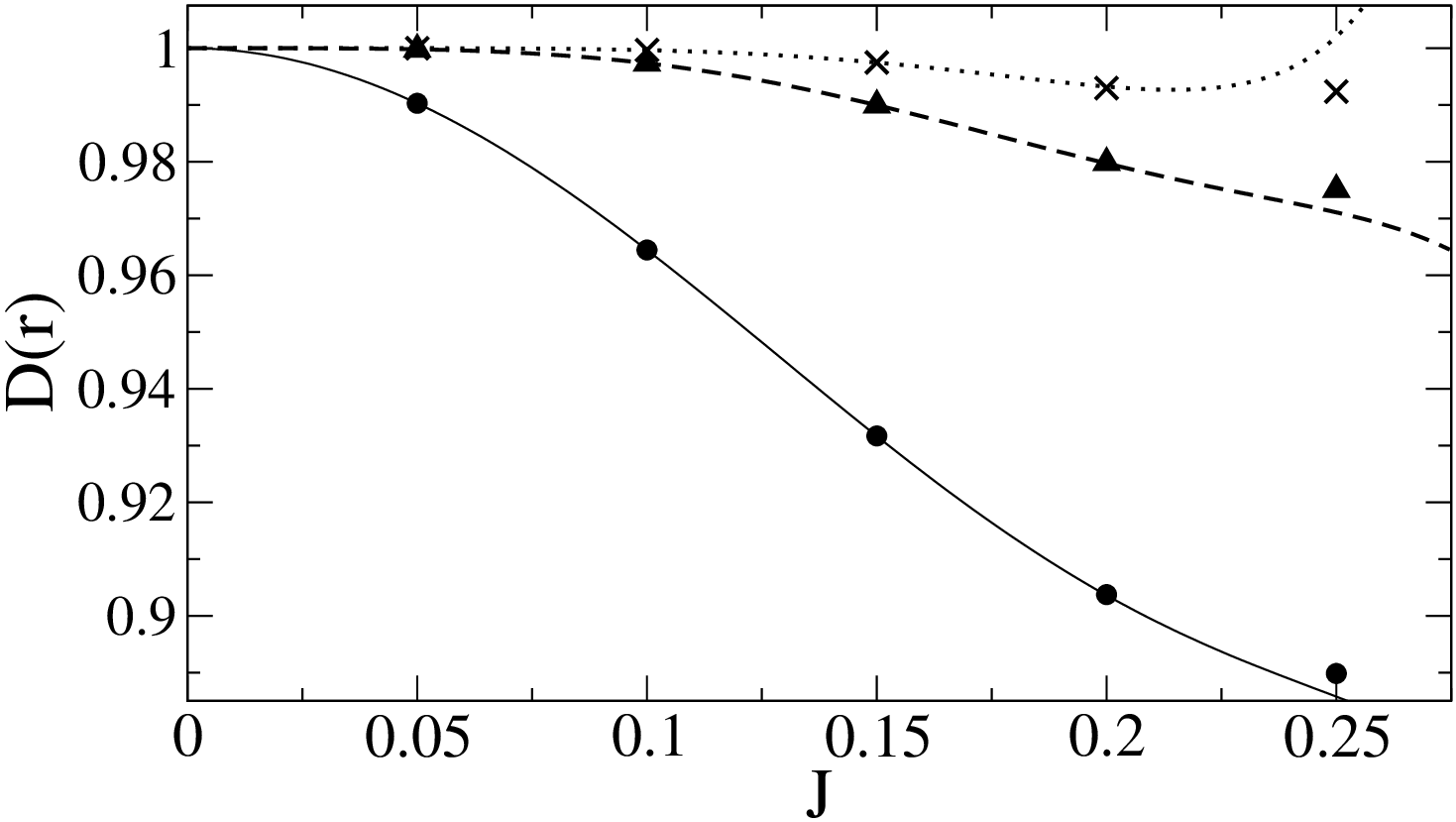}
\caption{Density-density correlations as a function of the tunneling $J$.
Numerics for $M=40$ system: $D(1)$ (dots), $D(2)$ (triangles), 
$D(3)$ (x's). Perturbative expansions  (\ref{D}):
solid, dashed and dotted lines represent $D(1)$, $D(2)$, $D(3)$, respectively. 
}
\label{fig4}
\end{figure}

The comparison between the expansion and the numerics is presented in 
Fig. \ref{fig4}: there is  a good agreement up to 
 $J$ equal to $0.2\sim0.225$, which may be a little surprising concerning 
high order of (\ref{D}).
One may attribute it  to presence  of open boundary conditions introducing 
small inhomogeneities in atoms density in our numerics.
On the other hand, the effects of open boundaries will always be present 
in an experimental setup proposed in  \cite{raizen}. Thus, our numerical 
calculations might actually better represent  the experimental situation
than the expansion (\ref{D}) derived for an idealized infinite lattice. 

Our perturbative predictions can be 
tested for self-consistency, e.g., the following  identity can be 
derived from the eigenequation for the ground state energy: 
$${\cal E}= -2 J  C(1)+ \frac{{\rm Var}(\hat{n})^2}{2}.$$
Using (\ref{energy_per_site}), (\ref{V}) and (\ref{K}) 
one easily shows that indeed it {\it exactly} holds up to ${\cal O}(J^{16})$ terms, as
expected from the orders of expansions presented in this paper.

Though it is  known that mean-field theory  works  badly 
in 1D, it is instructive at this point to compare our findings to  
predictions of the most popular mean-field, i.e., a Gutzwiller variational 
wave function \cite{jaksch,krauth}.
This approach predicts in 1D,  at unit filling factor, that the ground state 
is $|1,1,\dots\rangle$ for $0\le J\le1/11.6$, which is interpreted 
as a MI phase. It implies that in this $J$ range:  ${\cal E}=0$, 
${\rm Var}(\hat{n})=0$,
$C(r)=0$, $D(r)=1$. Our results analytically quantify the amount of discrepancy
between these mean-field predictions and the exact ones. It would be also 
instructive to compare our findings to  predictions of the 
perturbatively improved Gutzwiller approach \cite{schroll}.

\begin{widetext}
\begin{eqnarray}
D(1) &=& 1 - 4 J^2+ \frac{136}{3} J^4- \frac{2008}{27} J^6- \frac{150638}{81} J^8
	 + \frac{4897282}{729} J^{10} -\frac{415922848153}{14696640} J^{12}+
	 \frac{1022120948444278027}{7777461888000} J^{14},\nonumber\\
D(2) &=& 1 - \frac{100}{3} J^4+ \frac{2128}{3} J^6- 
          \frac{1156462}{243} J^8 -\frac{6848011}{729} J^{10} +
	  \frac{10808763042127}{44089920} J^{12}
	  -\frac{5150051155340205251}{3888730944000}J^{14},\nonumber\\
D(3) &=& 1 - \frac{13064}{27} J^6+ \frac{3727066}{243} J^8
	 - \frac{1588041877}{8748} J^{10} +
	 \frac{1710030328933}{2755620} J^{12}+
	 \frac{2208787916976404357}{370355328000}J^{14}.\nonumber\\
\label{D}
\end{eqnarray}
\end{widetext}

\section{Symmetry of perturbative expansions}
\label{secc}
Expansions of ${\cal E}$, $C(r)$, and $D(r)$
have well defined parity with respect to the $J\to-J$ transformation.
That is explained below, and provides an insight into the $J\to-J$ symmetry 
of the BHM. 
Additionally, it helps in determining the accuracy of our 
perturbative predictions. Indeed, the symmetry of perturbative 
expansions proven below 
implies that  if the last calculated nonvanishing term 
shows up in the $r$-th order, the expansion term in the 
$r+1$  order vanishes.

First, the ground state energy per site, Eq. (\ref{energy_per_site}), satisfies 
${\cal E}(J)={\cal E}(-J)$. 
To see it one 
performs the canonical transformation 
\begin{equation}
\hat{a}_i\to(-1)^i\hat{a}_i,
\label{dupa}
\end{equation}
which is a bosonic
equivalent of the Shiba transformation used to prove 
a similar symmetry property  of the Fermi-Hubbard model: Sec. 2.2.4
of \cite{FH}. After the transformation, Hamiltonian (\ref{H})  reads
$$\hat{H}= J \sum_{i=1}^M (\hat{a}_{i+1}^\dag \hat{a}_i + {\rm h.
c.}) +\frac{U}{2} \sum_{i=1}^M \hat{n}_i(\hat{n}_i-1),$$
iff (i) the system is infinite; 
(ii) the system consists  of even number of sites and periodic boundary 
conditions are applied; (iii) open boundary conditions are chosen.
To the end of this paper we assume that one of these conditions holds, and
that there exists a unique ground state.
The transformed Hamiltonian  is a $J\to-J$ version of (\ref{H}), 
so we get the desired prediction about the symmetry of ${\cal E}(J)$.

Second, we focus on 
$C(r)|_{J>0}=\langle+|\hat a_j^\dag\hat a_{j+r}|+\rangle$, where 
\begin{equation}
|+\rangle=\sum_{\{n_i\}} C_{\{n_i\}}(\hat{a}_1^\dag)^{n_1}\cdots(\hat{a}_M^\dag)^{n_M}
|{\rm vac}\rangle
\label{j}
\end{equation}
is a ground state of (\ref{H}) with $J>0$ [$C_{\{n_i\}}$ are some coefficients and
$\hat a_i|{\rm vac}\rangle=0$]. Using (\ref{dupa}), one gets that 
the ground state of the Hamiltonian (\ref{H}) with  $J<0$ is (up to a phase
factor): 
$$|-\rangle=\sum_{\{n_i\}} (-1)^{\sum j n_j} C_{\{n_i\}}
(\hat{a}_1^\dag)^{n_1}\cdots(\hat{a}_M^\dag)^{n_M}|{\rm vac}\rangle,$$
where $C_{\{n_i\}}$ are the same as in (\ref{j}).
An explicit calculation shows that 
$C(r)|_{J<0}=\langle-|\hat{a}_j^\dag\hat{a}_{j+r}|-\rangle= (-1)^r C(r)|_{J>0}$,
therefore   $C(r)\to (-1)^r C(r)$ as $J\to-J$: see (5) for an example
when $\sum n_i=M\to\infty$.

Third, following above  calculation, one 
shows that $D(r)$ and ${\rm Var}(\hat n)$ does not change under $J$ sign inversion.
The latter result is shadowed by explicit assumption of $J>0$ during Taylor expansion 
leading to (\ref{V}). 

Finally, we note that the ground state
of two component Bose-Einstein condensate in an optical lattice was found to
change after inversion of $J$ sign \cite{graham}. Our paper 
provides tools for a closer inspection of this interesting finding.

\section{Summary}
In this paper, we have derived and discussed  analytical
predictions on   important quantities describing physics of the
Mott phase: variance of on-site number operator, atom-atom and density-density  
correlation functions. We have also explored,  
for the first time to our knowledge, 
the fundamental $J\to-J$ symmetry of the BHM. 
We expect that our findings can be useful in interpretation of ongoing experiments 
performed in homogeneous lattices filled with ultracold atoms. 

We would like to acknowledge insightful discussions with 
Dominique Delande, Mark Raizen,  and Eddy  Timmermans,
as well as the financial support from 
the US Department of Energy. 
Calculations based on Vidal's algorithm were done in ICM UW under grant G29-10.  
J.Z. was supported by Polish Government scientific funds (2005-2008).

\appendix
\section{Perturbative expansion}
\label{pe}
Here we sketch some of  technical details of the nondegenerate 
perturbative expansions that we perform.

To introduce the notation: the Hamiltonian (\ref{H}) is re-written as 
$\hat{H}_0+\lambda\hat V$, where $\lambda=-J$, 
$\hat{H}_0= \frac{1}{2}\sum_{i=1}^M\hat{n}_i(\hat{n}_i-1)$,
$\hat V= \sum_{i=1}^M(\hat{a}_{i+1}^\dag \hat{a}_i + {\rm h. c.})$.

The ground state wave-function and eigenenergy up to a given order are 

\begin{equation}
|\Psi\rangle= \sum_{r=0}^S\lambda^r|\varphi(r)\rangle,
\ \ E=\sum_{r=0}^{S+1} \lambda^r {\cal E}_r,
\label{psi_E}
\end{equation}
\begin{equation}
|\varphi(r)\rangle= \sum_i a^{(r)}_i |i\rangle, \ \ \hat
H_0|i\rangle=e_i|i\rangle, \ \
\langle i|j\rangle=\delta_{ij}.
\label{ari}
\end{equation}
To keep the wave function (\ref{psi_E}) normalized to unity we 
must satisfy
$\langle\Psi|\Psi\rangle= 1+ {\cal O}(\lambda^{S+1})$. It 
implies  that  $\langle\varphi(0)|\varphi(0)\rangle= 1$ and \cite{granice} 
\begin{equation}
\langle\varphi(0)|\varphi(r)\rangle= -\frac{1}{2}\sum_{i=1}^{r-1}\langle
\varphi(i)|\varphi(r-i)\rangle, \ \ r=1,\dots,S.
\label{zero}
\end{equation}
The wave function (\ref{psi_E}) follows the eigen-equation
\begin{equation}
(\hat{H}_0+\lambda\hat{V})|\Psi\rangle=E|\Psi\rangle,
\label{eigen}
\end{equation}
which in the $\lambda^0$ order implies that 
$|\varphi(0)\rangle=a_0^{(0)}|0\rangle=|0\rangle$ (after proper choice of basis),
and ${\cal E}_0=e_0$.
In higher orders $a_0^{(r)}=\langle0|\varphi(r)\rangle$ 
is found from (\ref{zero}).

The coefficients $a^{(r\ge1)}_{i\neq0}$ (\ref{ari}) are obtained from projection 
of (\ref{eigen}) onto basis states $\{|i\neq0\rangle\}$ (\ref{ari})  
\begin{eqnarray}
a^{(r)}_i (e_i-e_0)+
\sum_j\langle i|\hat V|j\rangle a^{(r-1)}_j=\sum_{k=1}^{r-1}{\cal E}_k a^{(r-k)}_i.
\label{aaa}
\end{eqnarray}
The eigenenergies are found from a projection of (\ref{eigen}) onto
$|0\rangle$  \cite{granice}
\begin{eqnarray}
{\cal E}_r= \sum_j \langle0|\hat V|j\rangle a^{(r-1)}_j-
\sum_{k=1}^{r-1}{\cal E}_k a^{(r-k)}_0,
\label{eee}
\end{eqnarray}
where $r=1,\dots,S+1$.
Equations (\ref{zero}), (\ref{aaa}) and (\ref{eee}) have to be solved  
order by order starting from $r=1$. 

In our calculations 
\begin{equation}
|\varphi(0)\rangle:=|0\rangle=|1,1,\dots\rangle, 
\label{dup}
\end{equation}
and 
\begin{equation}
|i\rangle=|n_{1i},n_{2i},\dots,n_{Mi}\rangle, \ \ e_i=\frac{1}{2}\sum_k
(n_{ki}-1)n_{ki}.
\label{ini}
\end{equation}
Efficient manipulations on Fock states (\ref{ini}) are 
crucial for getting high order   expansions on a computer. 
We have achieved it by using  
a fast mapping $\{n_{1i},n_{2i},\dots,n_{Mi}\}\to i$
which was provided by the hash functions -- see \cite{knuth} for their general description.

Since $e_i$ (\ref{ini}) is an integer number and the 
action of $\hat V$ onto a Fock state produces factors of the form 
$\sqrt{n_{k\pm1}+1}\sqrt{n_k}$ ($n_k$ is the atom occupation of the k-th lattice  site), 
we see from (\ref{aaa}) and (\ref{eee}) that every $a_i^{(r)}$
coefficient will be in general equal to 
\begin{equation}
a^{(r)}_i=\sum_{s=1}^{s_{\rm max}}\frac{b\left(\{r,i\},s\right)}{c\left(\{r,i\},s\right)}
\sqrt{\prod_{k=1}^{k_{\rm max}} p\left(\{r,i\},s,k\right)},
\label{sy}
\end{equation}
where $b(\{r,i\},s)$ and $c(\{r,i\},s)$ are integer numbers and $p(\{r,i\},s,k)$ are prime
numbers satisfying $p(\{r,i\},s,k)\le r+1$ -- notice that when we develop the 
expansion around the Fock state (\ref{dup}) the highest on-site 
occupation in the $r$-th order is $r+1$ because it originates from 
$r$ consecutive actions of $\hat V$ operator onto (\ref{dup}). 
Similarly, one finds that $k_{\rm max}(\{r,i\},s)\le r$. 

We have done  the symbolic calculations by writing a proper ${\cal C}$ code.
Each set of $\{p(\{r,i\},s,k):k=1\dots k_{\rm max}\}$  has been encoded in
bits of a single integer number, while every $b(\{r,i\},s)/c(\{r,i\},s)$ ratio
has been allocated as a Gnu Multi Precision rational number \cite{gmp}.
The first choice minimizes the memory requirements on storage of
the prime number part, while the latter one allows for avoiding overflows 
occurring at high order expansions when $b(\{r,i\},s)$ and $c(\{r,i\},s)$ 
are allocated as integer  numbers.
To reduce  usage of memory the
$s_{\rm max}$ parameter has to be set as small as possible during allocation
process. By running the program we found that $s_{\rm max}=2$ is sufficient 
for all the calculations performed in this paper.

The results of the symbolic calculations were directly verified 
by calculation of the perturbative expansion in a numeric (as opposed to
a symbolic) way. There we treat $a^{(r)}_i$ as double precision numbers 
avoiding   complications of
 symbolic  manipulations. Results obtained numerically  
 are in agreement with the  symbolic ones within  accuracy provided by the 
double precision format.

\section{Noninteracting system}
\label{sf}
It is instructive to briefly summarize here the predictions for various
observables in the limit of negligible atom interactions, say 
$U\equiv0$ case. 

The $N$- particle ground state of the non-interacting $M$- site 
Hubbard model (\ref{H}) with periodic boundary conditions is given by 

\begin{equation}
|SF\rangle= \frac{1}{\sqrt{N!M^N}}\left(\sum_{i=1}^M\hat a_i^\dag\right)^N|{\rm vac}\rangle,
\label{gs}
\end{equation}
where $\hat a_i|{\rm vac}\rangle=0$ and $\langle SF|SF\rangle=1$ -- 
notice that (\ref{gs}) is nothing else than 
the $N$- particle zero momentum ground state \cite{stoof}. The ground state eigenenergy
equals $-2JN$.

After  some  algebra (\ref{gs}) can be rewritten to the form 
\begin{equation}
|SF\rangle=
\sqrt{\frac{N!}{M^N}}\sum_{\{n_i\}}\frac{|n_1,\dots,n_M\rangle}{\sqrt{n_1!\cdots n_M!}},
\label{gss}
\end{equation}
where the sum goes over all sets of boson on-site occupation (integer) 
numbers $\{n_i\}$ such that $\sum_i n_i=N$
and $0\le n_i\le N$. The wave-function (\ref{gss}) allows for an easy
calculation of different expectation values. Indeed, the summation 
can be expressed as $\sum_{\{n_i\}}=\sum_{n_1=0}^N\cdots\sum_{n_M=0}^N
\delta_{n_1+\cdots+n_M, N}$, then the Kronecker delta can be represented 
as $\delta_{l,m}=\frac{1}{2\pi}\int_0^{2\pi}d\varphi [\exp(i\varphi)]^{l-m}$, and finally
the integration over $\varphi$ can be replaced by the integration over $z=\exp(i\varphi)$
on the unit circle surrounding the origin of the complex $z$ plane. 
Usage of the residue theorem allows for getting that 
$$
{\rm Var}(\hat n)= \sqrt{\frac{N}{M}-\frac{N}{M^2}}, \ \ C(r)= \frac{N}{M}, \ \ D(r)=
\frac{N(N-1)}{M^2},
$$
for any integer $N>0$ and $M<r$. 

In this paper we are interested in the case of $N/M=1$ and $M\gg1$, i.e.,
\begin{equation}
{\rm Var}(\hat n)= 1, \ \ C(r)= 1, \ \  D(r)= 1.
\label{sf}
\end{equation}


\begin{thebibliography}{99}

\bibitem{jaksch} D. Jaksch, C. Bruder, J.I. Cirac, C.W. Gardiner, and P. Zoller,
Phys. Rev. Lett. {\bf 81}, 3108 (1998).

\bibitem{bloch}
M. Greiner, O. Mandel, T. Esslinger, T.W. H\"ansch, and I. Bloch,
Nature {\bf 415}, 39 (2002).

\bibitem{esslinger} T. St\"oferle, H. Moritz, C. Schori, M. K\"ohl, and T.
Esslinger, Phys. Rev. Lett. {\bf 92}, 130403 (2004). 

\bibitem{annals} D. Jaksch and P. Zoller, Ann. Phys. (N.Y.) {\bf 315}, 52
(2005).

\bibitem{hightc} 
W. Hofstetter, J.I. Cirac, P. Zoller, E. Demler, and M.D. Lukin, 
Phys. Rev. Lett. {\bf 89}, 220407 (2002).

\bibitem{sanpera} 
A. Sanpera, A. Kantian, L. Sanchez-Palencia, J. Zakrzewski, and M.
Lewenstein, Phys. Rev. Lett. {\bf 93}, 040401
(2004).

\bibitem{duan} L.M. Duan, E. Demler, and M.D. Lukin, Phys. Rev. Lett. {\bf
91},  090402 (2003).

\bibitem{bodzio1} 
B. Damski, H.-U. Everts,  A. Honecker, H. Fehrmann, L. Santos, and M.
Lewenstein, Phys. Rev. Lett. {\bf 95}, 060403 (2005).

\bibitem{fisher} M.P.A. Fisher, P.B. Weichman, G. Grinstein, and D.S.
Fisher, Phys. Rev. B {\bf 40}, 546 (1989).

\bibitem{sachdev} S. Sachdev, {\it Quantum Phase Transitions} 
(Cambridge University Press, Cambridge UK, 2001).

\bibitem{qc} 
D. Jaksch, H.-J. Briegel, J.I. Cirac, C.W. Gardiner, and P. Zoller, 
Phys. Rev. Lett. {\bf 82}, 1975 (1999);
	     G.K. Brennen, C.M. Caves, P.S. Jessen, and I.H. Deutsch,
	     Phys. Rev. Lett. {\bf 82}, 1060 (1999).

\bibitem{qmc} G.G. Batrouni, R.T. Scalettar, and G.T. Zimanyi, Phys. Rev.
Lett. {\bf 65}, 1765 (1990).

\bibitem{krauth} W. Krauth, M. Caffarel, and J.-P. Bouchaud, Phys. Rev. B
{\bf 45}, 3137 (1992).

\bibitem{dmrg} T.D. K\"uhner and H. Monien, Phys. Rev. B 
{\bf 58}, R14741 (1998); R.V. Pai, R. Pandit, H.R. Krishnamurthy, and
S. Ramasesha, Phys. Rev. Lett. {\bf 76}, 2937 (1996).

\bibitem{elstner} N. Elstner and H. Monien, Phys. Rev. B {\bf 59},
12184 (1999); cond-mat/9905367 (unpublished).

\bibitem{white} T.D. K\"uhner, S.R. White, and H. Monien, Phys. Rev. B
{\bf 61}, 12474 (2000).

\bibitem{batrouni} 
G.G. Batrouni {\it et al.}, Phys. Rev. 
Lett. {\bf 89}, 117203 (2002); G.G. Batrouni, F.F. Assaad, R.T. Scalettar, and
P.J.H. Denteneer, Phys. Rev. A {\bf 72}, 031601(R) (2005); 
P. Sengupta, M. Rigol, G.G. Batrouni, P.J.H. Denteneer, and R.T.
Scalettar, Phys. Rev. Lett. {\bf 95}, 220402 (2005).


\bibitem{raizen} T.P. Meyrath, F. Schreck, J.L. Hanssen,
C.-S. Chuu and M.G. Raizen,  Phys. Rev. A {\bf 71}, 041604(R) (2005);
M.G. Raizen, {\it private communication}.

\bibitem{osterloh} L. Amico, A. Osterloh, and F. Cataliotti, Phys. Rev. Lett.
                {\bf 95}, 063201 (2005).


\bibitem{Vidal} G. Vidal, Phys. Rev. Lett. {\bf 93}, 040502 (2004).

\bibitem{collath} A.J. Daley {\it et al.}, J. Stat. Mech.: Theor. Exp. P04005 (2004).


\bibitem{gelfand} M.P. Gelfand, R.R.P. Singh, and D.A. Huse, J. Stat. Phys. 
{\bf 59}, 1093 (1990).

\bibitem{spins} A. Honecker, Phys. Rev. B {\bf 59}, 6790 (1999); C. Knetter,
K.P. Schmidt, and G.S. Uhrig, Eur. Phys. J. B {\bf 36}, 525 (2003).

\bibitem{zin} E. Br\'ezin, J.C. Le Guillou, and J. Zinn-Justin, Phys. Rev. D
{\bf 15}, 1544 (1977).

\bibitem{zwerger} V.A. Kashurnikov, N.V. Prokof'ev, and B.V. Svistunov,
Phys. Rev. A {\bf 66}, 031601(R) (2002).

\bibitem{schroll} C. Schroll, F. Marquardt, and C. Bruder, Phys. Rev. A
{\bf 70}, 053609 (2004).

\bibitem{FH} 
F.H.L. Essler, H. Frahm, F. G\"ohmann, A. Kl\"umper, and V.E. Korepin, 
{\it The one dimensional Hubbard model} (Cambridge University Press, Cambridge UK, 2005).

\bibitem{graham} K.V. Krutitsky and R. Graham, Phys. Rev. Lett. {\bf 91},
240406 (2003).

\bibitem{granice} If the upper summation limit is smaller then the lower one,
the sum equals zero by definition.

\bibitem{knuth} D.E. Knuth, {\it The art of computer programming}
(Addison-Wesley, Boston USA, 1998) -- see volume 3 in the second edition.

\bibitem{gmp} www.swox.com/gmp.

\bibitem{stoof} D. van Oosten, P. van der Straten, and H.T.C. Stoof,
Phys. Rev. A {\bf 63}, 053601 (2001).

\end{thebibliography}
\end{document}